\documentclass[usenatbib,useAMS,usedcolumn]{mn2e}

\usepackage{epsfig,graphics,graphicx}
\usepackage{times}
\pdfoutput=1

\newcommand{\arcm}{\hbox{$^\prime$}}

\newcommand{\degree}{\hbox{$^\circ$}}
\newcommand{\rosat}{\emph{ROSAT}}
\newcommand{\chandra}{\emph{Chandra}}
\newcommand{\xmm}{\emph{XMM-Newton}}
\newcommand{\xmms}{\emph{XMM}}

\newcommand{\arcs}{\mbox{\arcm\arcm}}

\newcommand{\Zsol}{\ensuremath{Z_{\odot}}}

\newcommand{\Msol}{\ensuremath{M_{\odot}}}

\newcommand{\s}{\ensuremath{\mbox{~s}}}
\newcommand{\ps}{\ensuremath{\s^{-1}}}
\newcommand{\cm}{\ensuremath{\mbox{~cm}}}
\newcommand{\pcmsq}{\ensuremath{\cm^{-2}}}
\newcommand{\km}{\ensuremath{\mbox{~km}}}
\newcommand{\Mpc}{\ensuremath{\mbox{~Mpc}}}
\newcommand{\pMpc}{\ensuremath{\Mpc^{-1}}}
\newcommand{\kmpspMpc}{\ensuremath{\km \ps \pMpc\,}}
\newcommand{\erg}{\ensuremath{\mbox{~erg}}}
\newcommand{\ergps}{\ensuremath{\erg \ps}}
\newcommand{\ergpspcmsq}{\ensuremath{\erg \ps \pcmsq}}
\newcommand{\kmps}{\ensuremath{\km \ps}}

\newcommand{\Ho}{\ensuremath{H_\mathrm{0}}}

\newcommand{\Dtf}{\ensuremath{D_{\mathrm{25}}}}

\newcommand{\gtsim}{\,\rlap{\raise 0.5ex\hbox{$>$}}{\lower 1.0ex\hbox{$\sim$}}\,}

\newcommand{\nh}{\ensuremath{\mathrm{n}_\mathrm{H}}}
\begin{document}

\title[ 
The impact of sloshing in NGC~5044
] 
{ 
The impact of sloshing on the intra-group medium and old radio lobe of NGC~5044
}

\author[E. O'Sullivan et al.]  {Ewan O'Sullivan\footnotemark[1]$^{1}$, Laurence P. David$^{1}$, Jan
  M. Vrtilek$^{1}$\\
  $^{1}$ Harvard-Smithsonian Center for Astrophysics, 60 Garden Street, Cambridge, MA 02138 }

\date{Accepted 2013 October 7; Received 2013 October 2; in original form 2013 September 22}

\pagerange{\pageref{firstpage}--\pageref{lastpage}} \pubyear{2010}

\maketitle

\label{firstpage}

\begin{abstract} 
We present temperature and abundance maps of the central 125~kpc of the NGC~5044 galaxy group, based an a deep \xmm\ observation. The abundance map reveals an asymmetrical abundance structure, with the centroid of the highest abundance gas offset $\sim$22~kpc northwest of the galaxy centre, and moderate abundances extending almost twice as far to the southeast than in any other direction. The abundance distribution is closely correlated with two previously--identified cold fronts and an arc--shaped region of surface brightness excess, and it appears that sloshing, induced by a previous tidal encounter, has produced both the abundance and surface brightness features. Sloshing dominates the uplift of heavy elements from the group core on large scales, and we estimate that the southeast extension (the tail of the sloshing spiral) contains at least 1.2$\times$10$^5$\Msol\ more iron than would be expected of gas at its radius. Placing limits on the age of the encounter we find that if, as previously suggested, the disturbed spiral galaxy NGC~5054 was the perturber, it must have been moving supersonically when it transited the group core. We also examine the spectral properties of emission from the old, detached radio lobe southeast of NGC~5044, and find that they are consistent with a purely thermal origin, ruling out this structure as a significant source of spectrally hard inverse--Compton emission. 
\end{abstract}


\footnotetext[1]{E-mail: ejos@head.cfa.harvard.edu}

\section{Introduction}
\label{sec:intro}

The high throughput and angular resolution of the current generation of X--ray telescopes (\xmm\ and \chandra) has revealed a wealth of structure in the hot intergalactic medium of galaxy groups and clusters. This has facilitated the study of a number of astrophysical processes, most notably AGN feedback and mergers. One of the first classes of structures recognised in early \chandra\ observations were cold fronts, well-defined discontinuities in X-ray surface brightness in which the inner, brighter side of the front is cooler and denser \citep[e.g.,][]{Markevitchetal00,Vikhlininetal01}. Although cold fronts are often found in ongoing mergers, in many cases such fronts are observed in apparently relaxed groups and clusters, taking the form of arcs or spirals around the cool core \citep[e.g.,][]{MarkevitchVikhlinin07}. These are thought to be formed by the process of sloshing, in which the passage of a lower-mass subcluster or subgroup falling through the main cluster transfers angular momentum to the intra-cluster medium (ICM), setting it oscillating within the cluster potential \citep{Markevitchetal01}. This oscillation produces an expanding spiral pattern of discontinuities, as lower entropy gas from the cluster core is lifted outward, cooling adiabatically to maintain pressure equilibrium with its new surroundings. This process has been extensively modelled via numerical simulations \citep[e.g.,][]{AscasibarMarkevitch06,ZuHoneetal11,Roedigeretal11} and studied in detail in a number of groups and clusters \citep[see, e.g., references in][]{Roedigeretal11}. 

The cool gas transported out of the group or cluster core by the sloshing motions has typically been enriched by supernovae in the central dominant elliptical or cD galaxy, and these metals are also transported out to larger radii \citep[e.g.,][]{Simionescuetal10,dePlaaetal10,Roedigeretal11,Gastaldelloetal13,Ghizzardietal13,Canningetal13}. Merger--induced sloshing is therefore an important process for the redistribution of metals from galaxies out into the ICM, alongside uplift by the jets of group and cluster--central active galaxies \citep[e.g.,][]{Sandersetal04,Kirkpatricketal09,OSullivanetal11b}.

NGC~5044 is the X-ray brightest group in the sky, and at a redshift of $z$=0.00928, has an angular scale (1\arcs=185~pc) well-suited for the study of structures in its intra--group medium (IGM). The position of peak X--ray surface brightness agrees well with the optical centroid of the galaxy, but \rosat\ observations showed that despite the generally relaxed and regular appearance of the extended IGM, the galaxy is offset from the centroid of the large--scale X-ray emission, suggesting some degree of disturbance \citep{Davidetal94}. The \rosat\ data also showed that the cool core is asymmetrical, with a tail or plume of cool gas extending southeast. A joint \chandra\ and \xmms\ analysis found evidence of multiphase gas within the central 30~kpc, and identified a cold front northwest of the core \citep{Buoteetal03a}, and \citet{Gastaldelloetal09} later identified an inner cold front to the southeast. The presence of these two fronts, their relative distances from the galaxy centroid ($\sim$150\arcs\ and 350\arcs\ for the SE and NW fronts respectively) and the $\sim$150\kmps\ offset between NGC~5044 and the group mean velocity \citep{CelloneBuzzoni05,Mendeletal08} lead to the suggestion that the group is sloshing \citep{Gastaldelloetal09,Davidetal09}. 

Deeper \chandra\ observations revealed numerous small cavities in the IGM core, indicating that NGC~5044 has undergone several recent AGN outbursts \citep{Davidetal09,Davidetal11}. The isotropic distribution of the cavities suggests that gas motions within the core have produced a highly structured IGM with a wide range of abundances, mixed with bubbles of relativistic plasma inflated by the AGN jets. While gas motion in the core may be driven by the heating effects of the AGN, Giant Metrewave Radio Telescope (GMRT) observations at 235~MHz show the impact of the sloshing motions on larger scales; a one-sided jet undergoes two $\sim$90\degree\ bends around the inner SE cold front, which also abuts a detached radio lobe. Neither lobe or jet are detected at higher frequencies, indicating an ultra-steep spectral index (\gtsim 1.6) and probable great age. 

\citet{Davidetal11} consider the sloshing and suggest NGC~5054, a disturbed spiral galaxy $\sim$27\arcm\ SE of NGC~5044, as the likely perturber. A deep \xmm\ observation was analysed by \citet{Gastaldelloetal13} who identified a surface brightness excess east of the core, at larger radii than the cold fronts, and noted that this feature was predicted by sloshing simulations. They argue that the alignment of the fronts and excess suggests that our line of sight is almost parallel to the orbital plane of the perturber, and consistent with the perturber's current location being east of the group core. 

In this paper we use a deep, $\sim$100~ks \xmms\ observation to investigate
the temperature and abundance distributions of the IGM on large scales, and
the impact of sloshing on the metal distribution of the group. The paper is
organised as follows: In \textsection~\ref{sec:obs} we describe the \xmms\
observation, the data reduction and spectral extraction methods used in our
analysis; in \textsection~\ref{sec:maps} we present the temperature and
abundance maps of the group, examine the change in IGM properties across
the cold fronts, and examine the consistency of our results with \chandra\
observations; in \textsection~\ref{sec:lobe} we examine the X-ray emission
from the region of the detached radio lobe, and search for evidence of
non--thermal emission; finally, in \textsection~\ref{sec:discuss} we
discuss our results in the context of the enrichment of the IGM and the
impact of sloshing on the radio structures.

Throughout the paper we report uncertainties at the 68\% confidence limit. We adopt a cosmology with \Ho=70\kmpspMpc, $\Omega_{\rm m}$=0.3 and $\Omega_\Lambda$=0.7. Abundances were measured relative to the abundance table of \citet{GrevesseSauval98}.

\section{Observations and Data Reduction}
\label{sec:obs}
NGC~5044 was observed by \xmm\ on 2008 December 27 (ObsID 0554680101) for a
total of 128,819s. The EPIC instruments operated in full frame mode with
the thin optical blocking filter. The data were reduced and analysed using
the \xmms\ Science Analysis System (\textsc{sas v12.0.1}). Times when the
total count rate deviated from the mean by more than 3$\sigma$ were
excluded, leaving effective exposure times of 71~ks (EPIC-pn) and 98~ks
(MOS). Diffuse X-ray emission from the intra-group medium (IGM) of NGC~5044
fills the field of view of the EPIC instruments. This makes accurate
scaling and correction of blank-sky background data to match the
observation dataset difficult, and we therefore chose to model the
background, except when examining a region of limited size, where a local
background region could be used.

When modelling the background, we took two approaches. Where a limited
number of spectra were to be fitted (for example in radial profiles) we
used the \xmms-Extended Source Analysis Software (ESAS) and the general
model suggested by \citet{Snowdenetal04}. Point sources identified using
the CHEESE-BANDS task are excluded, and spectra and responses for each
region were extracted. Chip 5 of MOS 2 was found to be in an anomalous
state with mildly enhanced background, and was excluded from the ESAS
analysis.  An additional ROSAT All-Sky Survey (RASS) spectrum, extracted from an
annulus 1.75-2.25\degree\ ($\sim$1.16-1.5~Mpc) from the group centre using
the HEASARC X-ray Background
Tool\footnote{http://heasarc.gsfc.nasa.gov/cgi-bin/Tools/xraybg/xraybg.pl}
was also included to help constrain the soft X-ray background. All spectra
were fitted simultaneously in XSPEC 12.8.0 \citep{Arnaud96}. Energies
outside the range 0.3-10.0~keV (MOS) and 0.4-7.2~keV (pn) were ignored.

The particle component of the background was partially subtracted using
particle--only datasets scaled to match the event rates in regions of the
detectors which fall outside the field of view. Out--of--time (OOT) events
in the EPIC-pn data were statistically subtracted using scaled OOT spectra.
The remainder of the particle background was modelled with a powerlaw whose
index was linked across all annuli. As this element of the background is
not focused by the telescope mirrors, diagonal Ancillary Response Files
(ARFs) were used. The instrumental Al K$\alpha$ and Si K$\alpha$
fluorescence lines were modelled using Gaussians whose widths and energies
were linked across all annuli, but with independent normalisations. The
X--ray background was modelled with four components whose normalisations
were tied between annuli, scaling to a normalisation per square arcminute
as determined by the PROTON-SCALE task.  The cosmic hard X-ray background
was represented by an absorbed powerlaw with index fixed at $\Gamma$=1.46.
Thermal emission from the Galaxy, local hot bubble and/or heliosphere was
represented by one unabsorbed and two absorbed APEC thermal plasma models
with temperatures of 0.1, 0.1 and 0.25 keV respectively. The normalisations
of the APEC models were free to vary relative to one another. Absorption
was represented by the WABS model, fixed at the Galactic column density
\citep[\nh=4.87$\times$10$^{20}$\pcmsq, taken from the
Leiden/Argentine/Bonn survey,][]{Kalberlaetal05}. The RASS spectrum was
fitted using only the X-ray background components.

Since spectral extraction in ESAS is slow, and fitting is performed
simultaneously across multiple regions, this approach is unsuitable for
spectral mapping, where large numbers of spectra must be extracted and
fitted. We therefore adopted an alternative method when creating spectral
maps. We extracted spectra using standard SAS tools, removing bad pixels
and columns, and filtering the events lists to include only those events
with FLAG = 0 and patterns 0-12 (for the MOS cameras) or 0-4 (for the PN).
An out-of-time (OOT) events list was created and scaled OOT spectra were
used to statistically subtract the OOT contribution during analysis.

\begin{figure*}
\includegraphics[width=\textwidth,bb=36 295 576 496]{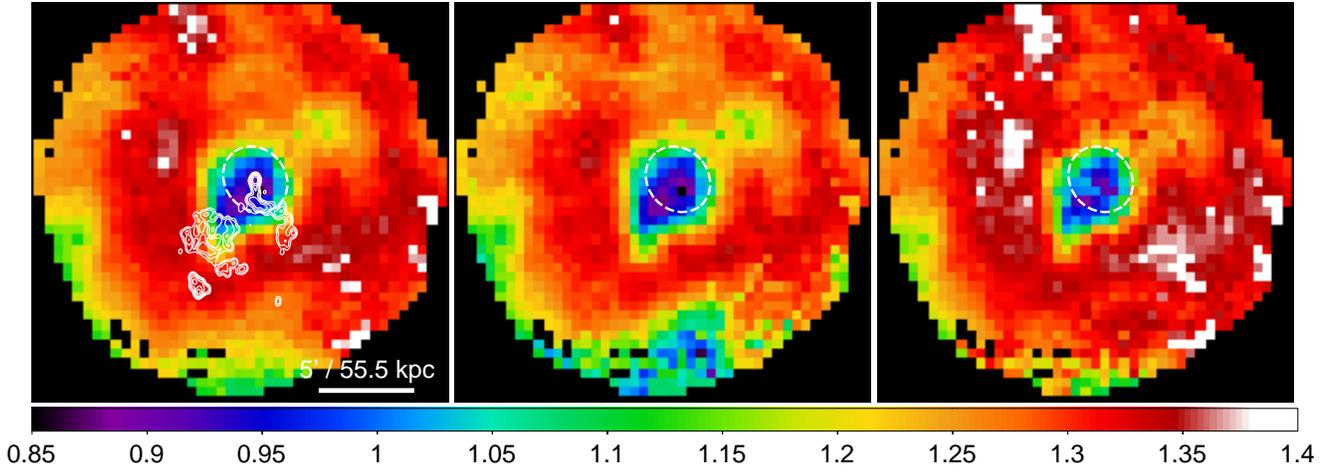}
\caption{\label{fig:Tmaps}\xmm\ projected temperature maps of NGC~5044, aligned with north up the page and west to the right. The left panel shows the best-fitting value, the centre and right panels the 1$\sigma$ lower and upper bounds respectively. The dashed ellipse shows the\Dtf\ contour of NGC~5044. GMRT 235~MHz contours are overlaid in the left panel, with half-power beam width (HPBW) =22\arcs$\times$16\arcs, the contour levels starting 3$\sigma$ above the r.m.s. noise (3$\sigma$=0.75~mJy~beam$^{-1}$) and increasing in steps of factor two.}
\end{figure*}

\begin{figure*}
\includegraphics[width=\textwidth,bb=36 295 576 496]{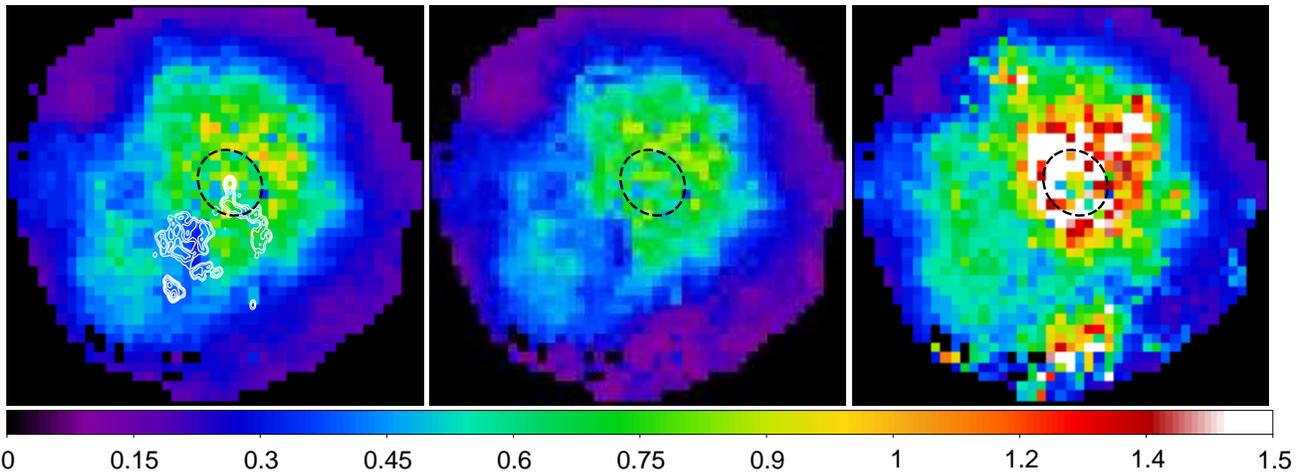}
\caption{\label{fig:Zmaps}\xmm\ projected abundance maps of NGC~5044, aligned with north up the page and west to the right. The left panel shows the best-fitting value, the centre and right panels the 1$\sigma$ lower and upper bounds respectively. The \Dtf\ ellipse and GMRT 235~MHz contours are overlaid, as described in figure~\ref{fig:Tmaps}.}
\end{figure*}

To reduce the computational time required to create spectra and responses,
we used a 8$\times$8 grid of Redistribution Matrix Files (RMFs) for each
camera, assigning each spectrum an RMF based on its central position.  We
also used the EVIGWEIGHT task to weight each event based on its position in
the detector, correcting for variations of effective area across the field
of view. This allows a single on-axis ARF to be used for each instrument.
As a result, the accuracy of the responses for each spectrum is slightly
reduced, but we consider this an acceptable trade--off given the decrease in
spectral extraction time and since features identified from the maps can
later be confirmed using standard responses. The spectra for each region
were then fitted using the same energy bands and essentially the same model
as used in the ESAS fitting described above. No background subtraction was
performed, so the powerlaw representing the particle background modelled
the entirety of this component. Fitting was carried out in CIAO
\textit{Sherpa} v4.5 \citep{Freemanetal01}.

\section{Spectral Maps}
\label{sec:maps}
In order to determine the large-scale two-dimensional projected distribution of
abundance and temperature in NGC~5044, we created spectral maps of the
system. We followed a process similar to that used for the \chandra\
observation \citep{Davidetal11} and in our previous \xmms\ studies of
groups \citep[e.g.,][]{OSullivanetal11b}. The field of view was divided
into a grid of 30\arcs-square pixels, whose centres correspond to the
centres of circular spectral extraction regions. The radius of these
circular regions was allowed to vary in steps of factor 1.4 between
15.9\arcs\ and 2\arcm, with the aim of including at least 6000 net
counts, summed over the three cameras. Since the background had not been
modelled at this stage, it was approximated using blank-sky data,
normalised to the observation data sets using the out of field of view
events in the 2-7 keV band. Regions whose surface brightness was too low to
provide 6000 net counts in the maximum extraction region were excluded from
further analysis. The spectral extraction regions can be larger than the pixel size, in which case neighbouring pixels will not be independent; in our \xmms\ maps pixels less than $\sim$5\arcm\ from the galaxy centroid are independent, but at larger radii the spectral extraction regions overlap. For this reason, and because of the nature of the background modelling approach used, the maps should be considered as tools for identifying regions of interest which can then be investigated and confirmed using standard spectral analysis techniques. Previous experience has shown that such maps are generally useful and reliable in this role \citep[e.g.,][]{OSullivanetal12,Davidetal11,OSullivanetal11a,OSullivanetal07}

\begin{figure*}
\includegraphics[width=\textwidth,bb=36 312 576 480]{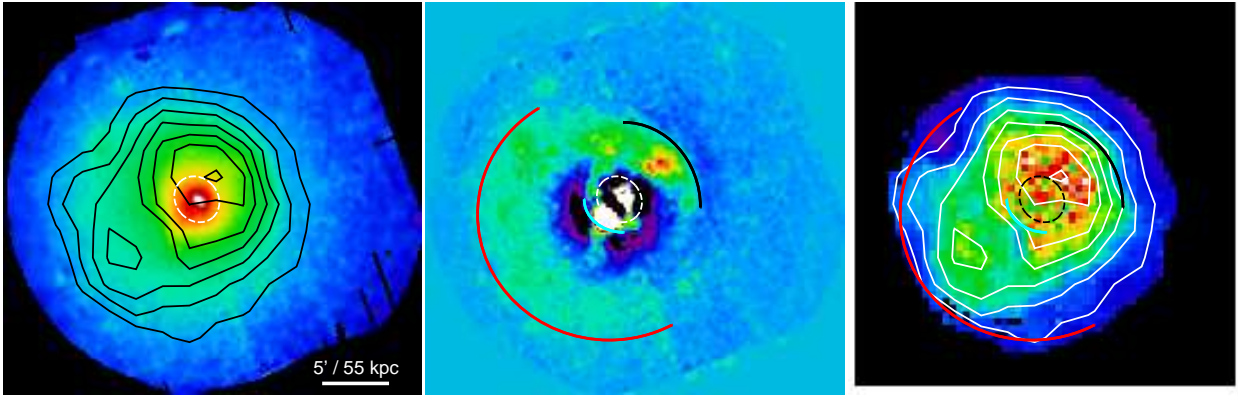}
\caption{\label{fig:SBvsZ}Comparison of X--ray surface brightness and abundance. The left panel shows a 0.3-2~keV MOS+pn image of NGC~5044, corrected for OOT events, with point sources removed and filled using the CIAO tool DMFILTH, and adaptively smoothed with a target signal--to--noise ratio of 20. The central panel shows the residuals after subtraction of a circular $\beta$-model, Gaussian smoothed with a 20\arcs\ kernel. The right hand panel shows the projected abundance map. The three panels have the same scale and orientation, with north toward the top of the page and west to the right. The approximate \Dtf\ contour of NGC~5044 is marked by a dashed ellipse, and solid arcs indicate the locations of the cold fronts (SE in cyan, NW in black) and surface brightness excess feature (in red) identified in previous studies \citep{Davidetal09,Gastaldelloetal09,Gastaldelloetal13}. Smoothed abundance contours starting at 0.2\Zsol\ and increasing in steps of 0.1\Zsol\ are overlaid on the left and right panels.}
\end{figure*}

Each spectrum was fitted using the background model described above, plus
an absorbed APEC thermal plasma component representing the IGM. The
resulting maps are shown in figures~\ref{fig:Tmaps} and \ref{fig:Zmaps}.
Fits typically have reduced $\chi^2$ values of 0.75-1.15. Uncertainties in
the map temperatures are generally small, $\sim$0.02~keV, equivalent to
$\sim$1\% of the measured value. The exception is a patch on the southern
edge of the map (most visible as a blue/green region in the centre panel of
Figure~\ref{fig:Tmaps}) where uncertainties rise to $\sim$10\%. This
corresponds to the location of MOS2 chip 5, suggesting that the larger
errors are caused by the enhanced background and altered background
spectrum in this chip. Abundance uncertainties are higher, typically
0.05-0.1\Zsol, with larger uncertainties in the core and southern patch. In
these regions, the 1$\sigma$ upper uncertainties can be $>$0.5\Zsol.
However, the structures seen in both maps are clearly visible in the
uncertainty maps.

The temperature map shows the extended, comma-shaped cool core (0.85-1~keV)
surrounded by higher temperature regions particularly to the south and
west. A trail of mid-temperature ($\sim$1.2~keV) material extends to the
northwest. The inner edge of this region was detected by \chandra\
\citep{Davidetal09,Davidetal11}. On larger scales the highest temperatures
($\sim$1.35~keV) persist to radii of $\sim$10\arcm (110~kpc) to the north and
west, but are less extended to the south and east, declining outside
$\sim$8\arcm (90~kpc).

The abundance map confirms the presence of a low abundance region
southeast of the core, coincident with the detached radio lobe seen at
235~MHz. This region is part of a band of mixed-abundance (0.3-0.45\Zsol)
emission extending around the southeast quadrant of the galaxy between 3
and 5\arcm\ (33-55~kpc) from the galaxy core, with a clump of somewhat
higher abundance material ($\sim$0.65\Zsol) beyond it. The main high
abundance region associated with the group core is offset to the northwest
of the galaxy, with the abundance centroid $\sim$2\arcm\ (22~kpc) from the
optical centroid. However, the upper abundance uncertainties in and around
the galaxy are large, probably owing to a combination of multi-temperature
gas along the line of sight, multi-phase gas within spectral extraction
regions, and uncertainties on the particle component of the background.
Fitting multi--temperature (and probably multi--abundance) emission with
single-temperature models will tend to bias the best--fit abundance low
\citep{Buotefabian98}. However, it is clear that the abundance distribution
is not centred on the galaxy, with the highest abundances offset to the
northwest, but moderate abundances most extended to the southeast.
Abundances decline to 0.1-0.2\Zsol\ outside 6-7\arcm\ on the northeast,
northwest and southwest sides, but remain enhanced out to $>$9.5\arcm\ in
the southeast.

Figure~\ref{fig:SBvsZ} shows a comparison between X-ray surface brightness
and the abundance map. The two cold fronts are marked; one $\sim$150\arcs\
(28~kpc) southeast of the galaxy centre, the other 350\arcs\ (65~kpc)
northwest. There is also the arc of excess emission discovered by
  \citet{Gastaldelloetal13} which extends from due west, inside the
northwest front, through north and around the east side of the galaxy. In
the southeast quadrant it lies between 6-11\arcm\ (66-122~kpc) from the
galaxy centre. The arc is a highly significant surface brightness
  feature. Comparing the surface brightness in 80\degree regions to
  northwest and southeast, we find that the surface brightness to the
  southeast exceeds that to the northwest by $\sim$45$\sigma$ between 400
  and 600\arcs \citep[see also Fig.~16 of][]{Gastaldelloetal13}. We find
a strong correlation between these features and the abundance distribution.
The highest abundances are found between the two cold fronts. The extended
moderate-metallicity gas to the southeast of the galaxy reaches a radius
very similar to the outer radius of the excess surface brightness arc,
though the morphology of the two features does not correspond exactly. The
moderate abundance material is more narrowly directed to the southeast than
the surface brightness excess, which extends to the northeast and south
into regions of low abundance. However, in general the abundance
distribution compares reasonably closely with surface brightness, at least
on arcminute scales, except in the core, where the lower temperatures and
high densities produce a brightness peak at the centre of the cool core.

Correlation between surface brightness and the temperature distribution is
less clear. Our map shows the moderate temperatures northwest of the galaxy
which form the outer cold front, but as noted by \citet{Davidetal11} the
tail of the cool core extends across the southeast cold front. As noted
previously, the southeast front lies just inside the detached radio lobe
seen at 235~MHz.

\subsection{Profiles across the cold fronts}

\begin{figure}
\includegraphics[width=\columnwidth,bb=36 266 576 525]{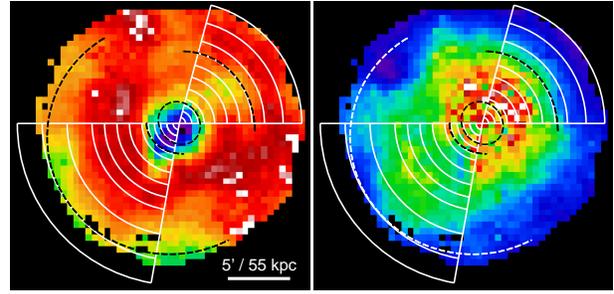}
\caption{\label{fig:angles}Regions used for extraction of the profiles shown in Figure~\ref{fig:prof}, overlaid on the projected temperature (left) and abundance (right) maps. Dashed arcs indicate the cold fronts and outer limit of the southeast surface brightness excess, and the dashed ellipse marks the \Dtf\ optical contour of NGC~5044.}
\end{figure}

\begin{figure}
\includegraphics[width=\columnwidth,bb=40 180 565 740]{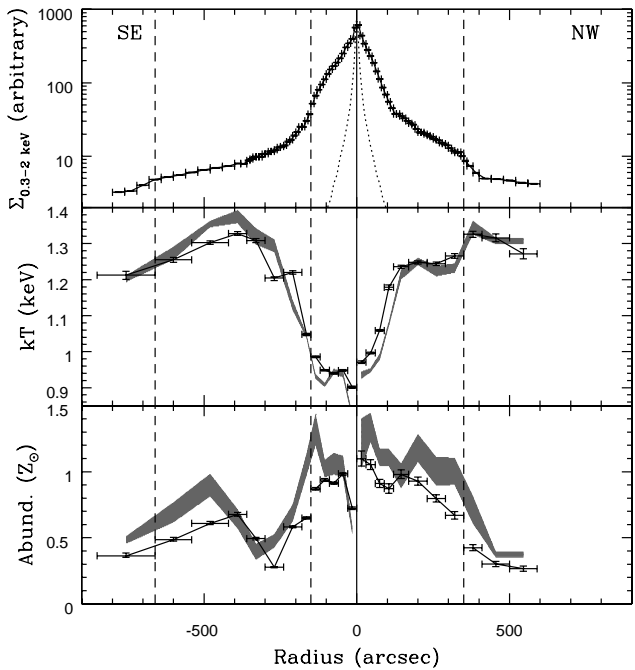}
\caption{\label{fig:prof}Profiles of projected 0.3-2~keV surface brightness, temperature and abundance extending northwest and southeast from the galaxy optical centroid, marked by the solid line at radius=0. Grey filled profiles mark the 1$\sigma$ error region for the deprojected temperature and abundance profiles. Dashed lines mark the positions of cold fronts and the outer edge of the SE surface brightness excess. The X--ray surface brightness profiles have been corrected for exposure but no background has been subtracted. The dotted lines indicate the scaled K--band optical surface brightness profile of NGC~5044.}
\end{figure}

To further examine the changes in IGM properties at the cold fronts and in the southeast surface brightness excess, we fitted spectral profiles extending from the galaxy optical centroid to the northwest and southeast. Spectra for the northwest profile were extracted between 270-345\degree\ (where 0\degree\ is due north) and for the southeast profile between 90-170\degree, with radii chosen to match those of the surface brightness features. Since these profiles have a limited number of spectra (11 and 13 per camera for the NW and SE profiles), we use the ESAS background subtraction and modelling formulism, including standard \xmms\ responses. The location of the profiles on the spectral maps is shown in Figure~\ref{fig:angles}, and the resulting profiles of temperature and abundance are shown in Figure~\ref{fig:prof}. Each partial annulus in the profiles contains \gtsim 20000 net counts. 

The profiles agree well with the maps, showing the greater extension of the
cool core and abundance profile to the southeast, and the more compact
abundance profile to the northwest, with a sharp decline at the cold front.
Despite the underlying declining trend in abundance with radius, it is
clear that the jump in abundance across both cold fronts is significant,
0.24$\pm$0.04\Zsol\ at the northwest front and 0.22$\pm$0.03\Zsol at the
southeast front. A temperature discontinuity is also visible across the
northwest front, with an increase of 0.06$\pm$0.01~keV outside a plateau of constant temperatures between 150-350\arcs. Both fronts are
visible as steepenings of the surface brightness profile, despite blurring
by the \xmms\ point spread function. Low abundances are found in the
innermost bin of the southeast profile, and at 250-300\arcs, the radius at
which the detached radio lobe is observed and the abundance map shows a
band of mixed--abundance gas. In both cases fitting a single--temperature
model to complex multi--phase gas is probably biasing the abundance
somewhat lower than the true value. The southeast abundance clump is
visible as a subsidiary peak in abundance values at $\sim$400\arcs,
associated with a flattening of the surface brightness profile. We note
that enhanced abundances extend southeast to the edge of the \xmms\ field
of view, $\sim$800\arcs\ ($\sim$150~kpc) from the galaxy centre, whereas
550\arcs\ (100~kpc) northwest of the galaxy abundance has declined to a
background level (0.27$\pm$0.02\Zsol\ compared to 0.38$\pm$0.01\Zsol).

Deprojected temperature and abundance profiles of the northwest and southeast profiles, fitted using the XSPEC PROJCT model, are also shown in Figure~\ref{fig:prof}. The deprojected profiles are similar to their projected counterparts, with somewhat higher abundances throughout, and a slightly larger cool core. It is notable that in the southeast profile the temperature in the core is lowest at the galaxy centroid (radius 0), then rises to $\sim$0.94~keV, then dips to 0.91~keV at $\sim$100\arcs. This behaviour is not visible in the projected profile, but is seen in the \xmms\ temperature map and the \chandra\ temperature map of \citet{Davidetal11}.

\begin{figure}
\includegraphics[width=\columnwidth,bb=40 210 565 740]{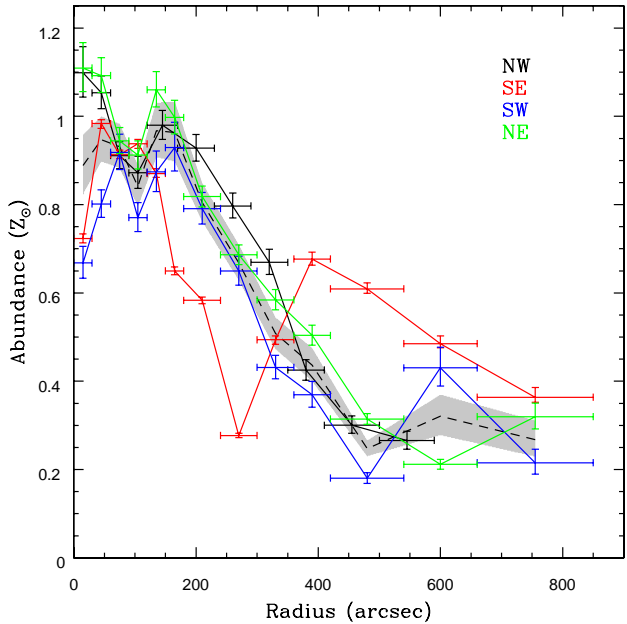}
\caption{\label{fig:4prof}Projected metallicity profiles in 80\degree\ sectors extending northwest (black), southeast (red), southwest (blue) and northeast (red). The NW and SE profiles are the same as those shown in Figure~\ref{fig:prof}. The grey hatched region and dashed line indicates the average abundance in the relatively undisturbed northeast and southwest sectors.}
\end{figure}

Figure~\ref{fig:4prof} shows the projected abundance profiles in the northwest and southeast quadrants, compared with equivalent profiles from 80\degree\ sectors to the northeast and southwest. Outside the core, the profiles from the (relatively undisturbed) NE and SW sectors agree reasonably well with one another in their general trend, though at radii beyond 300\arcs\ there is some divergence and increasing scatter between radial bins. Taking the mean of these two profiles as an approximation of the ``undisturbed'' pre--sloshing abundance profile, we see that the NW profile has enhanced abundances between 200-350\arcs, the region corresponding to the temperature plateau inside the NW surface brightness front. The SE abundance profile falls below the ``undisturbed'' mean profile out to $\sim$300\arcs, and is enhanced significantly above it at larger radii.

To test the impact of the adopted Galactic hydrogen column density on the profile fits, we refitted the southeast profile with the hydrogen column free to vary. The best fitting value, (5.70$\pm$0.12)$\times$10$^{20}$\pcmsq, is $\sim$17\% greater than the weighted mean column along the line of sight to NGC~5044 (as derived using the HEASOFT task NH) but is within the scatter of values found in the 1\degree$\times$1\degree\ region used to calculate the mean, and more similar to values found southeast of the group than to the northwest. The change in column has no impact on the shapes of the temperature or abundance profiles, and individual best-fit values generally change by less than their 1$\sigma$ uncertainty. We therefore conclude that the choice of hydrogen column is acceptable and is not affecting our results.

We also attempted to test the impact of multi--temperature gas on our fits,
since several studies have shown a requirement for multi--temperature
models when fitting spectra from the core of NGC~5044
\citep[e.g.,][]{Tamuraetal03,Buoteetal03a,Buoteetal03b,Davidetal09}.  Using
the previous, shorter ($\sim$19~ks) \xmms\ observation,
\citet{Buoteetal03a} showed that even when deprojected, the emission within
30~kpc of the galaxy centroid was best fitted with a two--temperature
model. \citet{Davidetal09} found a similar result using a deep \chandra\
observation, showing that fitting in different energy bands produces
different temperatures within 30~kpc but consistent temperatures at larger
radii. It seems likely that the need for multi--temperature models arises
in part from the temperature, abundance and density asymmetries of the IGM;
previous studies have generally used radial profiles azimuthally averaged
over the full 360\degree. Our deprojected northwest abundance profile is
similar to the azimuthally averaged two--temperature Fe abundances profiles
measured by \citet{Buoteetal03b} and falls significantly above their
single--temperature profile. This suggests our choice of limiting angles
for the profile reduces the degree of temperature variation in each
annulus, accounting for much of the effect which Buote et al.  required a
two--temperature model to correct.  However, \citet{Davidetal11} find
multi--temperature fits necessary even for small, homogeneous regions (as
do we, see \textsection~\ref{sec:lobe}) and it therefore seems likely that
genuinely multi--phase gas may be present.

Unfortunately, our radial profile spectra are unable to constrain a second source component in addition to the one--temperature deprojected source and background components. When a second thermal model is included in the fit, its temperature and/or normalisation either become unphysical or are so poorly constrained as to be meaningless. Even with parameters linked between groups of bins we are unable to find a meaningful fit. We do not see the strong residuals around the Fe-L complex which are characteristic of single--temperature fits to multi--temperature emission \citep[see e.g.,][Fig.~9]{Buoteetal03a}, supporting our view that our choice of regions has reduced the need for a more complex model. A thorough investigation of the impact of multi--phase gas which takes into account the asymmetries and discontinuities in the IGM probably requires a deeper observation. However, the fact that our fits do not require a second temperature component and agree reasonably well with the best previous abundance profiles leads us to conclude that they are reliable, at least outside the central 30~kpc.

\subsection{Comparison with \textit{Chandra}}

\begin{figure}
\includegraphics[width=\columnwidth,bb=36 241 576 550]{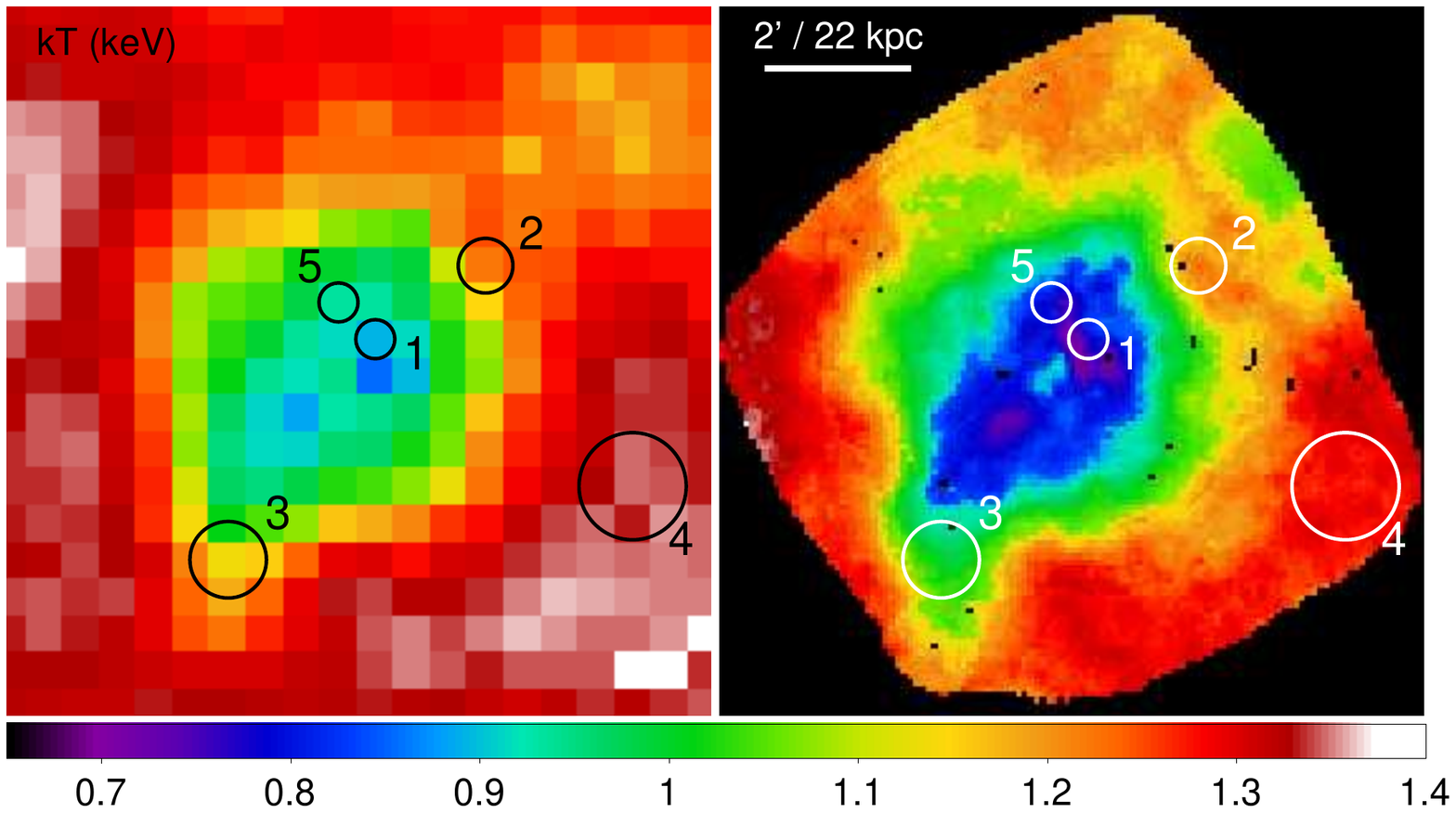}
\includegraphics[width=\columnwidth,bb=36 241 576 550]{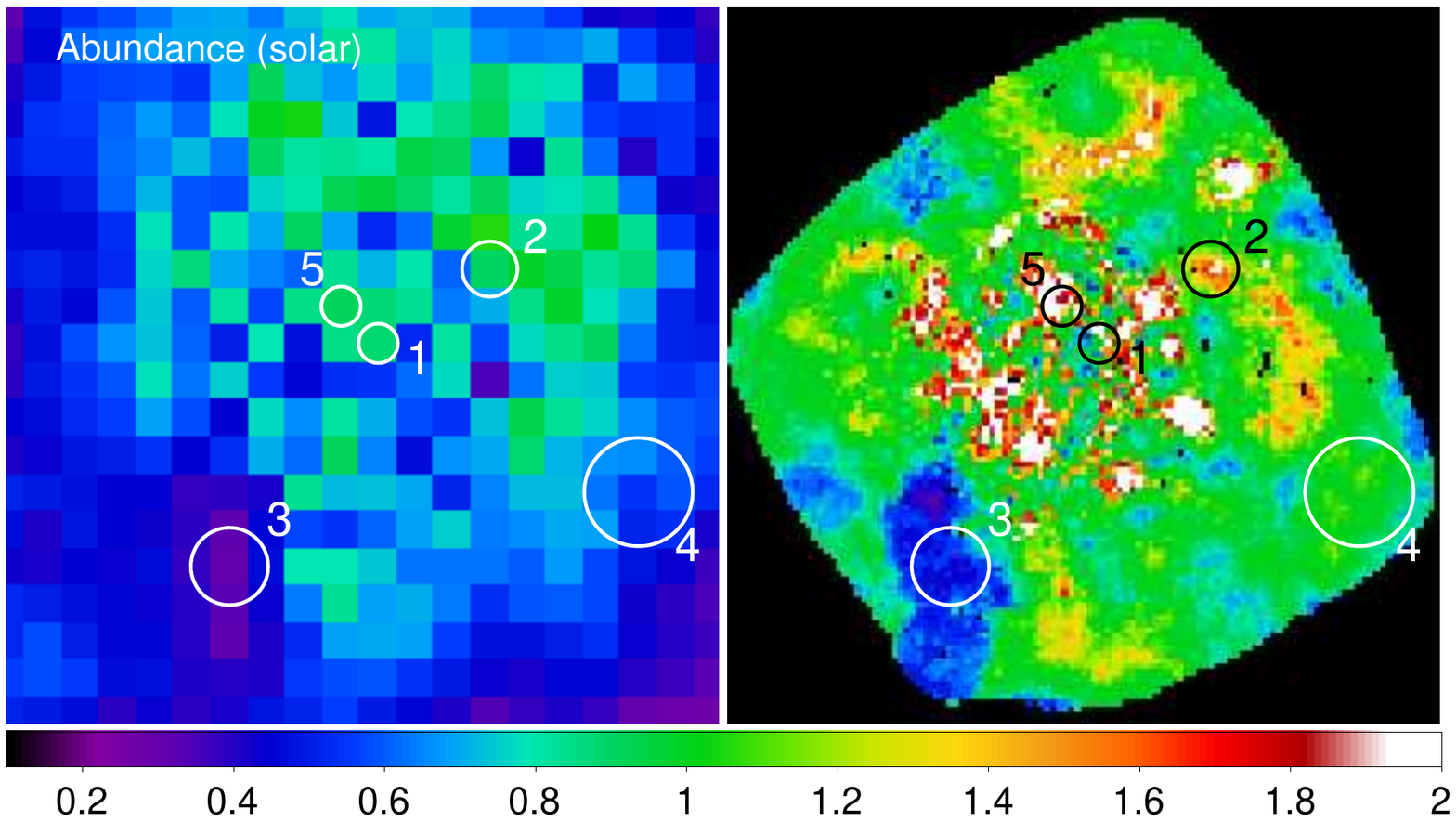}
\caption{\label{fig:XvsC}Comparison of \xmms\ (left) and \chandra\ ACIS-S3
  (right) maps of projected temperature (upper panels) and abundance (lower
  panels) in NGC~5044. The \chandra\ maps are taken \protect\citet{Davidetal11}, and have typical uncertainties of $\sim$3\% on temperature and $\sim$35\% on abundance. The results of spectral fits to the marked regions
  are listed in Table~\ref{tab:XvsC}. Orientation and angular scale of the
  four images is identical.  Note that the colour scales differ from those
  used elsewhere in this paper.}
\end{figure}

Comparing our maps with those of \citet{Davidetal11}, we find that while
the field of view and resolution of the \xmms\ and \chandra\ spectral maps
differ, several important structures are visible in both (see
Figure~\ref{fig:XvsC}). The most obvious are the comma-shaped cool core and
mid--temperature gas northwest of the core, and the low-abundance region
corresponding to the 235~MHz radio lobe (around region 3 in
Figure~\ref{fig:XvsC}). The \chandra\ and \xmms\ temperature maps also
agree well in terms of absolute values, the only significant differences
being found in the coolest parts of the cool core, where \chandra\ finds
temperatures 0.1-0.2~keV colder than those found by \xmms.

However, while the \xmms\ abundance map does show variations in metallicity within the core, it does not reproduce the clumps of super-solar abundances seen in the \chandra\ map. \chandra\ also finds somewhat higher abundances in the outer part of the core. Differences in background treatment (modelling for \xmms, blank-sky backgrounds for \chandra) seem unlikely to be responsible for this disagreement, given that both datasets are strongly source dominated in this region. Bias associated with multi-phase gas along the line of sight and within spectral extraction regions seems more likely. The \xmm\ spectral extraction regions used in this area range from 16-44\arcs\ in radius. The smallest regions are comparable in size to the \xmms\ point-spread function (PSF), with 16\arcs\ radius corresponding to a 70-75\% encircled energy fraction. In the same area, the \chandra\ maps used spectral extraction regions of radius $\sim$4-40\arcs, so in the brightest parts of the core, the \chandra\ map will be much less affected by mixing of spectra from regions of differing properties in the plane of the sky, though still affected by multi-temperature gas along the line of sight.  

To test the impact of the larger \xmms\ extraction regions, we selected five regions from the \xmms\ map, and extracted and fitted \chandra\ spectra using the \xmms\ extraction regions. Fitting was carried out in XSPEC v12.8.0, in the 0.5-5~keV energy band, using a hydrogen column of 4.94$\times$10$^{20}$\pcmsq, matching the values used by \citet{Davidetal11}. The regions used are shown in Figure~\ref{fig:XvsC} and the fit results are shown in Table~\ref{tab:XvsC}. In region 2-5 the \chandra\ and \xmms\ abundances agree well. In region 1, they agree only at the 2$\sigma$ level, despite large uncertainties on the \xmms\ abundance, but this region is in the coolest part of the cool core, with the most multi--temperature gas along the line of sight. Two temperature models with abundance linked between the two components provide acceptable fits to the \xmms\ data in region 1, 3 and 5, which lie in the cool core and its tail, but do not significantly alter the measured abundance. We therefore conclude that the disagreement between the \chandra\ and \xmms\ abundance maps is probably primarily caused by the different spectral extraction region sizes.

\begin{table*}
\begin{center}
\begin{tabular}{lcccccccccc}
Region & Extraction & \multicolumn{2}{c}{\xmms\ 1T} & \multicolumn{3}{c}{\xmms\ 2T} & \multicolumn{2}{c}{\chandra\ map} & \multicolumn{2}{c}{\chandra\ spectrum} \\
       & Radius  & kT & Abund. & kT$_h$ & kT$_c$ & Abund. & kT & Abund. & kT & Abund. \\
       & (\arcs) & (keV) & (\Zsol) & (keV) & (keV) & (\Zsol) & (keV) & (\Zsol) & (keV) & (\Zsol) \\
\hline\\[-3mm]
1 & 15.9 & 0.89$\pm$0.01 & 0.88$^{+0.02}_{-0.29}$ & 0.93$^{+0.02}_{-0.10}$ & 0.83$^{+0.60}_{-0.02}$ & 0.68$^{+0.23}_{-0.04}$ & 0.69-0.84 & 0.2-2.7 & 0.85$\pm$0.01 & 0.47$^{+0.02}_{-0.03}$ \\
2 & 22.3 & 1.22$^{+0.01}_{-0.02}$ & 0.90$^{+0.25}_{-0.06}$ & - & - & - & 1.12-1.23 & 1.0-2.4 & 1.29$^{+0.02}_{-0.01}$ & 1.11$^{+0.19}_{-0.15}$ \\
3 & 31.2 & 1.13$^{+0.02}_{-0.04}$ & 0.31$^{+0.46}_{-0.01}$ & 1.29$^{+0.02}_{-0.03}$ & 0.86$^{+0.03}_{-0.02}$ & 0.51$^{+0.48}_{-0.01}$ & 0.92-1.13 & 0.4-0.9 & 1.06$\pm$0.01 & 0.34$\pm$0.04 \\
4 & 43.7 & 1.35$^{+0.01}_{-0.03}$ & 0.54$^{+0.15}_{-0.05}$ & - & - & - & 1.25-1.30 & 0.7-1.2 & 1.33$^{+0.02}_{-0.01}$ & 0.71$\pm$0.09 \\
5 & 15.9 & 0.93$\pm$0.01 & 0.91$^{+0.03}_{-0.07}$ & 0.73-0.86 & 0.97$^{+0.06}_{-0.12}$ & 0.89$^{+0.11}_{-0.02}$ & 0.68$^{+0.69}_{-0.01}$ & 1.0-3.3 & 0.90$\pm$0.01 & 0.84$^{+0.08}_{-0.07}$ \\
\end{tabular}
\end{center}
\caption{\label{tab:XvsC}Temperatures and abundances measured from \xmms\ and \chandra\ for the five regions shown in Figure~\ref{fig:XvsC}. \xmm\ one--temperature (1-T)values are taken from the fits used in making the spectral maps, based on spectra extracted from circular regions whose radius is listed in column 2. Two--temperature (2-T) values use the same spectra and background models. \chandra\ map values indicate the range of temperatures and abundances found within these regions in the ACIS maps, whereas \chandra\ spectrum values are the results of models fitted to ACIS spectra extracted using these regions.}
\end{table*}

\section{The detached radio lobe}
\label{sec:lobe}
Based on spectral fits to \rosat\ and \textit{Rossi X-ray Timing Explorer}
PCA spectra, \citet{Henriksen11} claimed the detection of spectrally hard non-thermal emission from the NGC~5044 group, with a luminosity of $\sim$2.4$\times$10$^{42}$\ergps\ within 1\degree\  of the dominant galaxy ($\sim$670~kpc, using our chosen distance for luminosity and angular scale). Although their data did not have the spatial resolution to identify the origin of this emission, they suggested that the detached outer radio lobe
southeast of NGC~5044 was the most likely source, with the hard component arising from 
inverse--Compton scattering of cosmic microwave background photons by the
relativistic electron population. \citet{Davidetal11}
examined \chandra\ spectra from a much smaller, $\sim$15$\times$10~kpc elliptical region within the lobe and
confirmed that the gas in this region was likely multiphase, with a best
fit produced by a two-temperature absorbed VAPEC model with Fe and O free
to vary independently. The two temperatures, 0.71 and 1.23~keV, are
consistent with the range of temperatures observed in the \chandra\ and
\xmms\ spectral maps and are therefore likely to arise from the IGM
surrounding the lobe. The \chandra\ data showed no indication of any
high-temperature or non-thermal component arising from the contents of the
radio lobe, but are relatively shallow and insensitive to weak, spectrally
hard emission.

The deeper \xmm\ observation provides an opportunity to test again whether
any high-temperature component is present. We extracted spectra from a
circular region of radius 85\arcs ($\sim$16~kpc), chosen to approximate the
extent of the radio emission. We also extracted local background spectra
from a partial annulus with radii 110-185\arcs\ between angles of 25 and
250\degree\ (measured anti--clockwise from north). This region avoids the
tail or spur of cool emission extending from the core toward the lobe, and
should approximate the emission along the line of sight to the lobe.  To
allow a direct comparison with the \chandra\ results, these spectra were
initially fitted in the 0.5-7.0~keV range with absorption fixed at the
Galactic value.

We fitted the spectrum from the lobe with a 2-temperature VAPEC model,
initially with all elemental ratios fixed at the solar value, later freeing
O to compare with the \chandra\ results of \citet{Davidetal11}. Elemental
abundances were tied between the two temperature components. The resulting
fit parameters are shown in table~\ref{tab:lobe}.

\begin{table}
\begin{center}
\begin{tabular}{lcccc}
kT$_c$ & kT$_h$ & Fe & O & red. $\chi^2$/d.o.f.   \\
(keV) & (keV) & \Zsol\ & \Zsol\ &  \\
\hline\\[-3mm]
0.94$^{+0.02}_{-0.08}$ & 1.39$^{+0.12}_{-0.15}$ & 0.80$^{+0.12}_{-0.15}$ & - & 1.0873/803 \\
0.93$^{+0.16}_{-0.18}$ & 1.35$^{+0.12}_{-0.11}$ & 0.68$^{+0.07}_{-0.06}$ & 0.46$^{+0.14}_{-0.12}$ & 1.0831/802 \\
\end{tabular}
\end{center}
\caption{\label{tab:lobe} Two-temperature VAPEC fits to the detached lobe region. The upper line shows a fit with all elements fixed in solar ratios, the lower line a fit with Oxygen fitted separately from Fe. The temperature of the cooler component is given by kT$_c$, while kT$_h$ represents the hotter component.}
\end{table}

The temperatures in the lobe region are somewhat hotter than those found from the \chandra\ data, and when Fe and O are fitted separately the abundances found by both observatories are comparable, though poorly constrained (Fe=0.59-0.86 for \xmms, 0.71-1.00 for \chandra, O=0.31-0.79 for \xmms, 0.23-0.98 for \chandra). Some of these differences may arise from the different \chandra\ and \xmms\ extraction regions, and from our use of a local background where David et al. used a blank--sky background. 

To search for a spectrally hard non--thermal emission component, we tested single--temperature and two--temperature thermal models with an additional powerlaw component. We fit the MOS and pn spectra in the 0.3-10~keV and 0.4-7.2~keV bands respectively to take advantage of the high energy sensitivity of \xmms. We found that a single--temperature--plus--powerlaw model produces a significantly worse fit than the two--temperature thermal model (reduced $\chi^2$=1.1156 for 817 degrees of freedom, compared to 1.0667), with a powerlaw index significantly flatter than that found by Henriksen ($\Gamma$=1.56$^{+0.31}_{-0.45}$ compared to Henriksen's 2.64-2.82). The flux from the powerlaw is $\sim$3.7$\times$10$^{-14}$\ergpspcmsq\ in the 0.5-15~keV band used by Henriksen, a factor \textbf{$\sim$365} less than their measurement for the group as a whole. Fitting a two--temperature model with a powerlaw, we found that the best-fitting powerlaw normalisation always approximated zero and the powerlaw index is always poorly constrained; the spectrally soft thermal components account for essentially all the emission detected by \xmms. The 1$\sigma$ upper limit on the powerlaw flux fixing $\Gamma$=2.7, approximating Henriksen's best fit value, is $\sim$3$\times$10$^{-15}$\ergpspcmsq.

Taking the flux of the powerlaw in the two--temperature model as an
upper limit on the inverse--Compton flux of the lobe, we can place a lower
limit on the equipartition magnetic field $B_{eq}$, based on the ratio of
inverse--Compton and synchrotron fluxes \citep{GovoniFeretti04}. We measure
a 235~MHz flux of 117~mJy at 235~MHz from the GMRT image and adopt the
lower limit of $\alpha\geq$1.6 for the spectral index of the lobe at radio
frequencies \citep{Davidetal09}. From these values, we estimate the
magnetic field to be $B_{eq}$\gtsim 3.8~$\mu$G. This is considerably lower
that the estimate of $B_{eq}$=23~$\mu$G found by \citet{Davidetal09} on the
basis of pressure equilibrium between the lobe and the surrounding IGM, but
this is unsurprising since we were unable to detect a powerlaw component in
our two--temperature model. A lower inverse--Compton flux or steeper
spectral index would imply a stronger magnetic field. If we adopt
$B_{eq}$=23~$\mu$G as the true magnetic field strength in the lobe, we find
that the expected 0.5-15~keV inverse--Compton flux is
$\sim$2.7$\times$10$^{-17}$\ergpspcmsq\ (or
L$_{0.5-15~keV}$=5.1$\times$10$^{36}$\ergps).

In summary, while these results do not rule out inverse--Compton emission from the lobe, they strongly suggest that it is very weak compared to the thermal emission in the \xmms\ band.

\section{Discussion and Conclusions}
\label{sec:discuss}

It is clear from the abundance map that the sloshing of the NGC~5044 IGM has strongly influenced the distribution of metals. Indeed, although there is a correlation between radio structures, cavities, and abundance in the group core, indicating that nuclear jet activity has affected the abundance distribution, \citep{Davidetal09,Davidetal11}, on larger scales sloshing is much more important than (recent) AGN outbursts in determining the abundance distribution. Numerical modelling of interacting clusters has shown that much of the enriched material transported out from the core to larger radii by sloshing motions will remain at those radii, permanently broadening the abundance distribution \citep{Roedigeretal11}. It therefore appears that at present, sloshing is the dominant mechanism for transporting metals produced by the stellar population of NGC~5044 out to enrich the IGM, at least on scales $\sim$30-150~kpc.

\begin{figure}
\includegraphics[width=\columnwidth,bb=0 0 628 680]{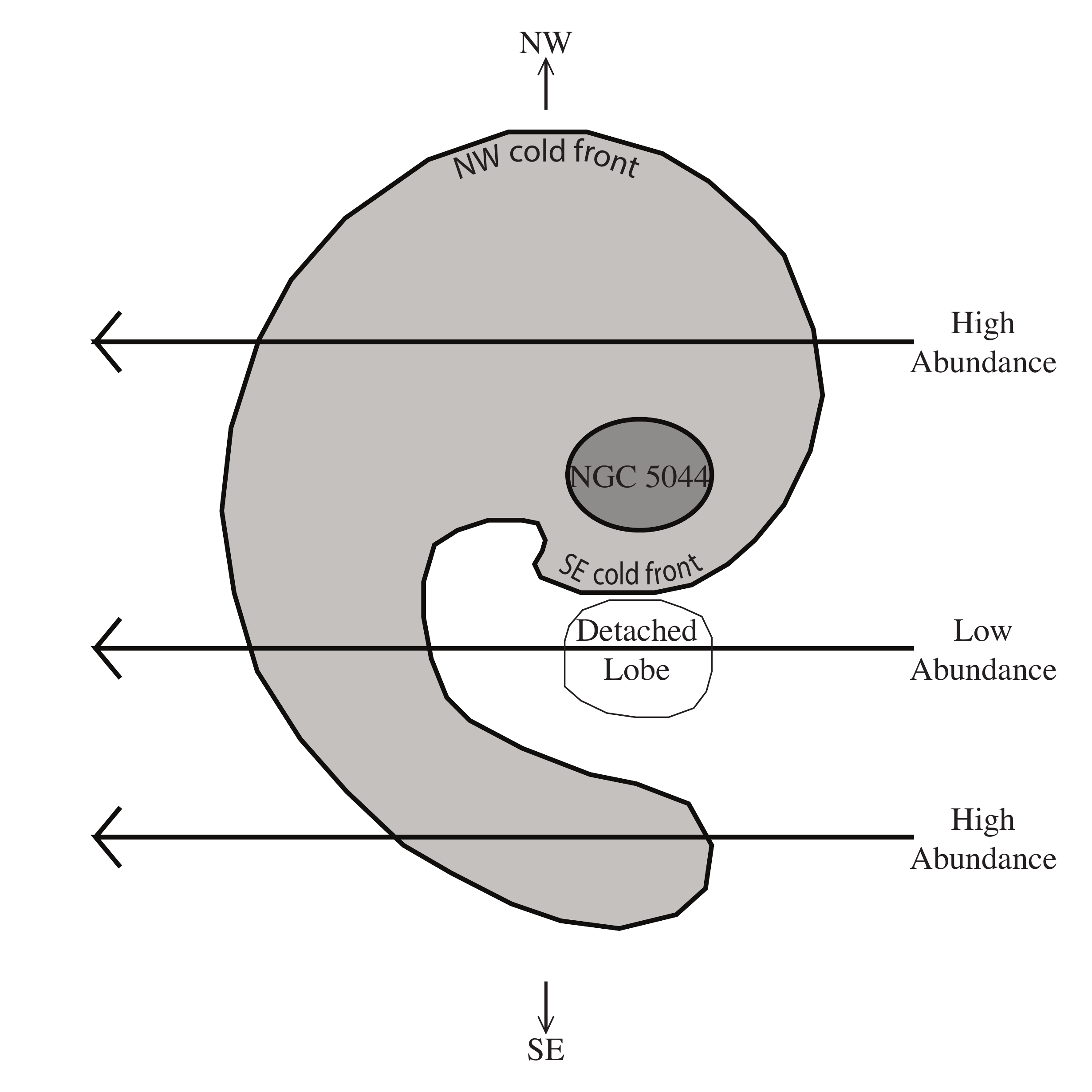}
\caption{\label{fig:diagram}Schematic diagram of the distribution of enriched gas in NGC~5044. Northwest is toward the top of the page, southeast to the bottom, northeast out of the page and southwest into the page. The plane of the sky is perpendicular to the page and the large horizontal arrows, which represent lines of sight from the observer. The light grey region is the enriched ``galaxy'' gas, surrounded by low metallicity IGM gas (white) and enclosing the NGC~5044 itself (dark grey). Different lines of sight pass through differing amounts of enriched gas producing the projected abundance pattern we observe, with high abundance around the galaxy and in the tail, and low abundance along the line of sight passing through the detached lobe. Note the the diagram is not to scale.}
\end{figure}

The lack of any spiral pattern in abundance or temperature, and the
presence of secondary abundance clump separated from the core by a lower
abundance band, confirms that the orbital plane of perturber is
approximately parallel to our line of sight. If the orbital plane were
closer to the plane of the sky (perpendicular to our line of sight), we
would expect to see a cool, enriched spiral of gas extending out from the
core, with a hotter, low abundance spiral extending inward from the IGM.
However, while the surface brightness residuals could be considered to
contain spiral features, it is clear that the abundance map does not. We
must therefore be viewing the sloshing spiral edge-on, with the arm of
enriched gas initially extending along the line of sight, curving through
the plane of the sky and then arcing back into the line of sight. Our sight
lines can be considered as passing through combinations of cool, enriched
``galaxy'' gas, and hotter, low--abundance IGM gas. The line of sight
through the galaxy passes through both the abundance peak and the base of
the uplifted enriched sloshing spiral, combining the highest abundance gas
and the greatest depth of enriched gas along the line of sight. The
medium-metallicity band in which the 235~MHz detached lobe is found looks
through the shortest depth of enriched gas, as the arm of the sloshing
spiral is there aligned close to the plane of the sky. The higher abundance
clump to the southeast is the tail of the enriched spiral, extended along
our line of sight, with a greater depth of enriched gas.
Figure~\ref{fig:diagram} illustrates these lines of sight through the
group.

From the deprojection analysis of the southeast radial profile shown in
Figure~\ref{fig:prof} we can estimate the density profile, the approximate
gas mass and (assuming that Fe emission dominates the abundance
measurements) iron mass in each radial bin. The spiral distribution of the
sloshing gas is not accounted for by the deprojection, which assumes
spherical symmetry and simple radial variations in temperature and
abundances, but for our purposes the relatively minor inaccuracies
introduced are acceptable. Comparing the differences between the southeast
abundance profile with the ``undisturbed'' mean of the northeast and
southwest profiles, and excluding the inner four bins where
multi--temperature gas in the core is likely to bias the abundances, we
estimate that there is a deficit of $\sim$3.7$\times$10$^4$\Msol\ of iron
between 125-325\arcs, and an excess of $\sim$1.2$\times$10$^5$\Msol\
between 325-650\arcs. We ignore the outermost bin of the profile since its
density is not deprojected and the gas mass would thus be overestimated.
These iron masses emphasise that in a sloshing system enriched gas is
uplifted from the core; the iron deficit at moderate radii is insufficient
to account for the excess at large radii. The sloshing spiral is much
broader at small radii than the sector used for spectral extraction, and
probably extends further than our profile; these masses are therefore
probably underestimates. Nonetheless, sloshing has uplifted a significant
mass of enriched gas to at least twice its previous radius.

Precisely estimating the age of the sloshing structures would require a
tailored simulation of the NGC~5044 system.  \citet{Roedigeretal12} note
that simulations of sloshing in clusters suggest that sloshing fronts
expand outward at a relatively constant rate of $\sim$55~kpc~Gyr$^{-1}$, at
least for low--mass systems. If this expansion rate holds for NGC~5044, it
would suggest ages of 1.2~Gyr and 500~Myr for the two cold fronts. However,
the $\sim$150\kmps\ offset of NGC~5044 from the group mean velocity exceeds
this expansion rate ($\sim$54\kmps) by a factor $\sim$3, suggesting that in
this case it may be an underestimate.  We can estimate (very approximate)
ages for the cold fronts based on the local sound speed $v_s$, assuming the
fronts expand at no more than 0.5$\times$$v_s$. For the outer NW front, the
local sound speed of $\sim$450\kmps\ suggests a travel time from the group
core of 280~Myr. The equivalent values for the inner SE front are 400\kmps\
and 140~Myr.

Taking the age of the outer cold front as an indicator of the time since the passage of the perturber through the group core, we can estimate that for a front age of 1.2~Gyr, if NGC~5054 is the cause of the sloshing, it must have a velocity of $\sim$240\kmps\ in the plane of the sky. Alternatively, if the front is only 280~Myr old, the plane--of--sky velocity of NGC~5054 would be $\sim$1050\kmps. The recession velocity of NGC~5054 is 1075\kmps\ less than that of NGC~5044 \citep{Mendeletal08} suggesting that in either case it would have been highly supersonic when passing through the core of NGC~5044. 

It is also notable that since the inner cold front appears to have
distorted the radio structures visible at 235~MHz, its age may provide a
lower limit on the time since the AGN outburst responsible for their
formation. 140-500~Myr is an exceptional age for such a small--scale
structure, which in a relaxed IGM would be expected to have buoyantly risen
out of the core, expanding and fading with time; the sound crossing time
between the AGN and lobe is only 20~Myr \citep{Davidetal09}.  However, its
location between the SE cold front and SE surface brightness excess, in the
mixed--abundance band, suggests an explanation of how it has remained
(apparently) close to the core. The lobe is probably encapsulated in the
arm of hotter, low-abundance IGM gas which has been drawn in toward the
group core by the sloshing motion, and the effect of these large--scale IGM
motions has exceeded the buoyant forces which would tend to move the lobe
outward. This also explains the strong bending of the 235~MHz jet or
filament, which twists around the cold front, probably crossing between the
two spirals of enriched outflowing gas, and metal--poor inflowing IGM gas.
Increases in pressure associated with these motions or the impingement of
the SE cold front onto the inner edge of the lobe may have caused
compression of its magnetic field and reacceleration of its particle
population, leading to a enhancement of its radio brightness.
Unfortunately, confirming the age of the lobe via radio measurements is
probably unfeasible. Given the estimated magnetic field of 23~$\mu$G
\citep{Davidetal09}, and conservatively assuming no energy losses from
adiabatic expansion or inverse-Compton scattering, we would expect the
break in the spectral index of the radio emission to be at (or below)
$\sim$0.8-10~MHz, below the observable waveband of current observatories.

\medskip
\noindent{\textbf{Acknowledgements}}\\
We thank the anonymous referee for a rapid and thorough reading of the paper.
This work is based on observations obtained with XMM-Newton, an ESA science
mission with instruments and contributions directly funded by ESA Member
States and NASA. Support for the analysis was provided by the National
Aeronautics and Space Administration through the Astrophysical Data
Analysis programme, award NNX13AE71G.

\bibliographystyle{mn2e}
\bibliography{../paper}

\begin{thebibliography}{}

\bibitem[\protect\citeauthoryear{{Arnaud}}{{Arnaud}}{1996}]{Arnaud96}
{Arnaud} K.~A.,  1996, in {Jacoby} G.~H.,  {Barnes} J.,  eds, Astronomical Data
  Analysis Software and Systems V Vol.~101 of Astronomical Society of the
  Pacific Conference Series, {XSPEC: The First Ten Years}.
p.~17

\bibitem[\protect\citeauthoryear{{Ascasibar} \& {Markevitch}}{{Ascasibar} \&
  {Markevitch}}{2006}]{AscasibarMarkevitch06}
{Ascasibar} Y.,  {Markevitch} M.,  2006, ApJ, 650, 102

\bibitem[\protect\citeauthoryear{Buote \& Fabian}{Buote \&
  Fabian}{1998}]{Buotefabian98}
Buote D.,  Fabian A.,  1998, MNRAS, 296, 977

\bibitem[\protect\citeauthoryear{{Buote}, {Lewis}, {Brighenti} \&
  Mathews}{{Buote} et~al.}{2003a}]{Buoteetal03a}
{Buote} D.~A.,  {Lewis} A.~D.,  {Brighenti} F.,    Mathews W.~G.,  2003a, ApJ,
  594, 741

\bibitem[\protect\citeauthoryear{{Buote}, {Lewis}, {Brighenti} \&
  Mathews}{{Buote} et~al.}{2003b}]{Buoteetal03b}
{Buote} D.~A.,  {Lewis} A.~D.,  {Brighenti} F.,    Mathews W.~G.,  2003b, ApJ,
  595, 151

\bibitem[\protect\citeauthoryear{{Canning}, {Sun}, {Sanders}, {Clarke},
  {Fabian}, {Giacintucci}, {Lal}, {Werner}, {Allen}, {Donahue}, {Johnstone},
  {Nulsen} \& {Sarazin}}{{Canning} et~al.}{2013}]{Canningetal13}
{Canning} R.~E.~A.,  {Sun} M.,  {Sanders} J.~S.,  {Clarke} T.~E.,  {Fabian}
  A.~C.,  {Giacintucci} S.,  {Lal} D.~V.,  {Werner} N.,  {Allen} S.~W.,
  {Donahue} M.,  {Johnstone} R.~M.,  {Nulsen} P.~E.~J.,    {Sarazin} C.~L.,
  2013, arXiv:1305.0050

\bibitem[\protect\citeauthoryear{{Cellone} \& {Buzzoni}}{{Cellone} \&
  {Buzzoni}}{2005}]{CelloneBuzzoni05}
{Cellone} S.~A.,  {Buzzoni} A.,  2005, MNRAS, 356, 41

\bibitem[\protect\citeauthoryear{{David}, {Jones}, {Forman} \&
  {Daines}}{{David} et~al.}{1994}]{Davidetal94}
{David} L.~P.,  {Jones} C.,  {Forman} W.,    {Daines} S.,  1994, ApJ, 428, 544

\bibitem[\protect\citeauthoryear{{David}, {Jones}, {Forman}, {Nulsen},
  {Vrtilek}, {O'Sullivan}, {Giacintucci} \& {Raychaudhury}}{{David}
  et~al.}{2009}]{Davidetal09}
{David} L.~P.,  {Jones} C.,  {Forman} W.,  {Nulsen} P.,  {Vrtilek} J.,
  {O'Sullivan} E.,  {Giacintucci} S.,    {Raychaudhury} S.,  2009, ApJ, 705,
  624

\bibitem[\protect\citeauthoryear{{David}, {O'Sullivan}, {Jones}, {Giacintucci},
  {Vrtilek}, {Raychaudhury}, {Nulsen}, {Forman}, {Sun} \& {Donahue}}{{David}
  et~al.}{2011}]{Davidetal11}
{David} L.~P.,  {O'Sullivan} E.,  {Jones} C.,  {Giacintucci} S.,  {Vrtilek} J.,
   {Raychaudhury} S.,  {Nulsen} P.~E.~J.,  {Forman} W.,  {Sun} M.,    {Donahue}
  M.,  2011, ApJ, 728, 162

\bibitem[\protect\citeauthoryear{{de Plaa}, {Werner}, {Simionescu}, {Kaastra},
  {Grange} \& {Vink}}{{de Plaa} et~al.}{2010}]{dePlaaetal10}
{de Plaa} J.,  {Werner} N.,  {Simionescu} A.,  {Kaastra} J.~S.,  {Grange}
  Y.~G.,    {Vink} J.,  2010, A\&A, 523, A81

\bibitem[\protect\citeauthoryear{{Freeman}, {Doe} \& {Siemiginowska}}{{Freeman}
  et~al.}{2001}]{Freemanetal01}
{Freeman} P.,  {Doe} S.,    {Siemiginowska} A.,  2001, in {J.-L.~Starck \&
  F.~D.~Murtagh} ed., Society of Photo-Optical Instrumentation Engineers (SPIE)
  Conference Series Vol.~4477 of Society of Photo-Optical Instrumentation
  Engineers (SPIE) Conference Series, {Sherpa: a mission-independent data
  analysis application}.
p.~76

\bibitem[\protect\citeauthoryear{{Gastaldello}, {Buote}, {Temi}, {Brighenti},
  {Mathews} \& {Ettori}}{{Gastaldello} et~al.}{2009}]{Gastaldelloetal09}
{Gastaldello} F.,  {Buote} D.~A.,  {Temi} P.,  {Brighenti} F.,  {Mathews}
  W.~G.,    {Ettori} S.,  2009, ApJ, 693, 43

\bibitem[\protect\citeauthoryear{{Gastaldello}, {Di Gesu}, {Ghizzardi},
  {Giacintucci}, {Girardi}, {Roediger}, {Rossetti}, {Brighenti}, {Buote},
  {Eckert}, {Ettori}, {Humphrey} \& {Mathews}}{{Gastaldello}
  et~al.}{2013}]{Gastaldelloetal13}
{Gastaldello} F.,  {Di Gesu} L.,  {Ghizzardi} S.,  {Giacintucci} S.,  {Girardi}
  M.,  {Roediger} E.,  {Rossetti} M.,  {Brighenti} F.,  {Buote} D.~A.,
  {Eckert} D.,  {Ettori} S.,  {Humphrey} P.~J.,    {Mathews} W.~G.,  2013, ApJ,
  770, 56

\bibitem[\protect\citeauthoryear{{Ghizzardi}, {De Grandi} \&
  {Molendi}}{{Ghizzardi} et~al.}{2013}]{Ghizzardietal13}
{Ghizzardi} S.,  {De Grandi} S.,    {Molendi} S.,  2013, Astron.~Nachr., 334,
  422

\bibitem[\protect\citeauthoryear{{Govoni} \& {Feretti}}{{Govoni} \&
  {Feretti}}{2004}]{GovoniFeretti04}
{Govoni} F.,  {Feretti} L.,  2004, International Journal of Modern Physics D,
  13, 1549

\bibitem[\protect\citeauthoryear{{Grevesse} \& {Sauval}}{{Grevesse} \&
  {Sauval}}{1998}]{GrevesseSauval98}
{Grevesse} N.,  {Sauval} A.~J.,  1998, Space Sci.~Rev., 85, 161

\bibitem[\protect\citeauthoryear{{Henriksen}}{{Henriksen}}{2011}]{Henriksen11}
{Henriksen} M.~J.,  2011, ApJ, 726, 9

\bibitem[\protect\citeauthoryear{{Kalberla}, {Burton}, {Hartmann}, {Arnal},
  {Bajaja}, {Morras} \& {P{\"o}ppel}}{{Kalberla} et~al.}{2005}]{Kalberlaetal05}
{Kalberla} P.~M.~W.,  {Burton} W.~B.,  {Hartmann} D.,  {Arnal} E.~M.,  {Bajaja}
  E.,  {Morras} R.,    {P{\"o}ppel} W.~G.~L.,  2005, A\&A, 440, 775

\bibitem[\protect\citeauthoryear{{Kirkpatrick}, {Gitti}, {Cavagnolo},
  {McNamara}, {David}, {Nulsen} \& {Wise}}{{Kirkpatrick}
  et~al.}{2009}]{Kirkpatricketal09}
{Kirkpatrick} C.~C.,  {Gitti} M.,  {Cavagnolo} K.~W.,  {McNamara} B.~R.,
  {David} L.~P.,  {Nulsen} P.~E.~J.,    {Wise} M.~W.,  2009, ApJ, 707, L69

\bibitem[\protect\citeauthoryear{{Markevitch}, {Ponman}, {Nulsen}, {Bautz},
  {Burke}, {David}, {Davis}, {Donnelly}, {Forman}, {Jones}, {Kaastra},
  {Kellogg}, {Kim} \& {et al.}}{{Markevitch} et~al.}{2000}]{Markevitchetal00}
{Markevitch} M.,  {Ponman} T.~J.,  {Nulsen} P.~E.~J.,  {Bautz} M.~W.,  {Burke}
  D.~J.,  {David} L.~P.,  {Davis} D.,  {Donnelly} R.~H.,  {Forman} W.~R.,
  {Jones} C.,  {Kaastra} J.,  {Kellogg} E.,  {Kim} D.-W.,    {et al.} 2000,
  ApJ, 541, 542

\bibitem[\protect\citeauthoryear{{Markevitch} \& {Vikhlinin}}{{Markevitch} \&
  {Vikhlinin}}{2007}]{MarkevitchVikhlinin07}
{Markevitch} M.,  {Vikhlinin} A.,  2007, Phys. Rep., 443, 1

\bibitem[\protect\citeauthoryear{{Markevitch}, {Vikhlinin} \&
  {Mazzotta}}{{Markevitch} et~al.}{2001}]{Markevitchetal01}
{Markevitch} M.,  {Vikhlinin} A.,    {Mazzotta} P.,  2001, ApJ, 562, L153

\bibitem[\protect\citeauthoryear{{Mendel}, {Proctor}, {Forbes} \&
  {Brough}}{{Mendel} et~al.}{2008}]{Mendeletal08}
{Mendel} J.~T.,  {Proctor} R.~N.,  {Forbes} D.~A.,    {Brough} S.,  2008,
  MNRAS, 389, 749

\bibitem[\protect\citeauthoryear{{O'Sullivan}, {Giacintucci}, {Babul},
  {Raychaudhury}, {Venturi}, {Bildfell}, {Mahdavi}, {Oonk}, {Murray},
  {Hoekstra} \& {Donahue}}{{O'Sullivan} et~al.}{2012}]{OSullivanetal12}
{O'Sullivan} E.,  {Giacintucci} S.,  {Babul} A.,  {Raychaudhury} S.,  {Venturi}
  T.,  {Bildfell} C.,  {Mahdavi} A.,  {Oonk} J.~B.~R.,  {Murray} N.,
  {Hoekstra} H.,    {Donahue} M.,  2012, MNRAS, 424, 2971

\bibitem[\protect\citeauthoryear{{O'Sullivan}, {Giacintucci}, {David}, {Gitti},
  {Vrtilek}, {Raychaudhury} \& {Ponman}}{{O'Sullivan}
  et~al.}{2011}]{OSullivanetal11b}
{O'Sullivan} E.,  {Giacintucci} S.,  {David} L.~P.,  {Gitti} M.,  {Vrtilek}
  J.~M.,  {Raychaudhury} S.,    {Ponman} T.~J.,  2011, ArXiv e-prints, 735, 11

\bibitem[\protect\citeauthoryear{{O'Sullivan}, {Giacintucci}, {David},
  {Vrtilek} \& {Raychaudhury}}{{O'Sullivan} et~al.}{2011}]{OSullivanetal11a}
{O'Sullivan} E.,  {Giacintucci} S.,  {David} L.~P.,  {Vrtilek} J.~M.,
  {Raychaudhury} S.,  2011, MNRAS, 411, 1833

\bibitem[\protect\citeauthoryear{{O'Sullivan}, {Vrtilek}, {Harris} \&
  {Ponman}}{{O'Sullivan} et~al.}{2007}]{OSullivanetal07}
{O'Sullivan} E.,  {Vrtilek} J.~M.,  {Harris} D.~E.,    {Ponman} T.~J.,  2007,
  ApJ, 658, 299

\bibitem[\protect\citeauthoryear{{Roediger}, {Br{\"u}ggen}, {Simionescu},
  {B{\"o}hringer}, {Churazov} \& {Forman}}{{Roediger}
  et~al.}{2011}]{Roedigeretal11}
{Roediger} E.,  {Br{\"u}ggen} M.,  {Simionescu} A.,  {B{\"o}hringer} H.,
  {Churazov} E.,    {Forman} W.~R.,  2011, MNRAS, 413, 2057

\bibitem[\protect\citeauthoryear{{Roediger}, {Kraft}, {Machacek}, {Forman},
  {Nulsen}, {Jones} \& {Murray}}{{Roediger} et~al.}{2012}]{Roedigeretal12}
{Roediger} E.,  {Kraft} R.~P.,  {Machacek} M.~E.,  {Forman} W.~R.,  {Nulsen}
  P.~E.~J.,  {Jones} C.,    {Murray} S.~S.,  2012, ApJ, 754, 147

\bibitem[\protect\citeauthoryear{{Sanders}, {Fabian}, {Allen} \&
  {Schmidt}}{{Sanders} et~al.}{2004}]{Sandersetal04}
{Sanders} J.~S.,  {Fabian} A.~C.,  {Allen} S.~W.,    {Schmidt} R.~W.,  2004,
  MNRAS, 349, 952

\bibitem[\protect\citeauthoryear{{Simionescu}, {Werner}, {Forman}, {Miller},
  {Takei}, {B{\"o}hringer}, {Churazov} \& {Nulsen}}{{Simionescu}
  et~al.}{2010}]{Simionescuetal10}
{Simionescu} A.,  {Werner} N.,  {Forman} W.~R.,  {Miller} E.~D.,  {Takei} Y.,
  {B{\"o}hringer} H.,  {Churazov} E.,    {Nulsen} P.~E.~J.,  2010, MNRAS, 405,
  91

\bibitem[\protect\citeauthoryear{{Snowden}, {Collier} \& {Kuntz}}{{Snowden}
  et~al.}{2004}]{Snowdenetal04}
{Snowden} S.~L.,  {Collier} M.~R.,    {Kuntz} K.~D.,  2004, ApJ, 610, 1182

\bibitem[\protect\citeauthoryear{{Tamura}, {Kaastra}, {Makishima} \&
  {Takahashi}}{{Tamura} et~al.}{2003}]{Tamuraetal03}
{Tamura} T.,  {Kaastra} J.~S.,  {Makishima} K.,    {Takahashi} I.,  2003, A\&A,
  399, 497

\bibitem[\protect\citeauthoryear{{Vikhlinin}, {Markevitch} \&
  {Murray}}{{Vikhlinin} et~al.}{2001}]{Vikhlininetal01}
{Vikhlinin} A.,  {Markevitch} M.,    {Murray} S.~S.,  2001, ApJ, 551, 160

\bibitem[\protect\citeauthoryear{{ZuHone}, {Markevitch} \& {Lee}}{{ZuHone}
  et~al.}{2011}]{ZuHoneetal11}
{ZuHone} J.~A.,  {Markevitch} M.,    {Lee} D.,  2011, ApJ, 743, 16

\end{thebibliography}

\label{lastpage}
\end{document}